\documentclass[11pt]{article}
\pdfoutput=1

\usepackage{a4wide}
\usepackage[fleqn]{amsmath}
\usepackage{hyperref}

\usepackage{fourier} 
\usepackage[scaled=0.875]{helvet} 

\RequirePackage{amsmath,amssymb,amsxtra,amsthm}

\RequirePackage{mathrsfs}

\RequirePackage{setspace}
\RequirePackage{array}
\RequirePackage{booktabs}

\RequirePackage{braket}

\RequirePackage[pdftex,final]{graphicx}
\graphicspath{{Plots/}}

\RequirePackage{colortbl}
\RequirePackage{xcolor}
\colorlet{titlerowcolor}{gray!15}

\usepackage{cite}

\newcommand{\be}{\begin{equation}}
\newcommand{\ee}{\end{equation}}
\newcommand{\bea}{\begin{eqnarray}}
\newcommand{\eea}{\end{eqnarray}}

\newcommand{\cN}{\mathcal{N}}

\numberwithin{equation}{section}
\numberwithin{table}{section}
\numberwithin{figure}{section}

\author{
  \begin{minipage}{0.97\linewidth}
    \vspace{1cm}
    \begin{center}
      \begin{small}
        \textbf{Carlo Angelantonj} $^{1}$, \textbf{Ignatios Antoniadis}$^{2,3}$ and  \textbf{Marine Samsonyan}$^{4}$
     \end{small}
    \end{center}
    \vspace{.3cm} \hspace{1.3cm}\begin{minipage}{.75\linewidth}
      {\it \begin{footnotesize}
          \begin{itemize}
          \item[${}^1$] Dipartimento di Fisica, Universit\`a di Torino, and INFN Sezione di Torino
          \\
            Via Pietro Giuria 1, 10125 Torino, Italy
          \item[${}^2$] LPTHE, CNRS UMR 7589,
         Sorbonne Universit\'es and UPMC  Paris 6
         \\4 place Jussieu,  75252 Paris cedex 05, France
            \item[${}^3$] Albert Einstein Center for Fundamental Physics, ITP
            \\
            University of Bern, Sidlerstrasse 5, 3012 Bern, Switzerland
            \item[${}^4$] Theoretical Physics Department, CERN, CH-1211 Geneva 23, Switzerland           
\end{itemize}
        \end{footnotesize}}
    \end{minipage}
    \vspace{1cm}
  \end{minipage}
}

\date{}

\title{\vspace{3cm}
  \begin{huge}
    \textbf{A string realisation of  $\varOmega$-deformed \\ Abelian $\cN =2^*$ theory}
  \end{huge}
}

\begin{document}

\begin{titlepage}
  \maketitle
  \thispagestyle{empty}

  \vspace{-14cm}
  \begin{flushright}
   \end{flushright}

 \begin{flushright}
 CERN-TH-2017-038
\end{flushright} 

  \vspace{11cm}

  \begin{center}
    \textsc{Abstract}\\
  \end{center}

The $\mathscr{N}=2^*$ supersymmetric gauge theory is a massive deformation of $\mathscr{N}=4$, in which the adjoint hypermultiplet gets a mass. We present a D-brane realisation of the (non-)Abelian $\mathscr{N}=2^*$ theory, and compute suitable topological amplitudes, which are expressed as a double series expansion. The coefficients determine couplings of higher-dimensional operators in the effective supergravity action that involve powers of the anti-self-dual $\mathscr{N}=2$ chiral Weyl superfield and of self-dual gauge field strengths superpartners of the D5-brane coupling modulus. In the field theory limit, the result reproduces the Nekrasov partition function in the two-parameter $\varOmega$-background, in agreement with a recent proposal.
 
\vfill

{\small
\begin{itemize}
\item[E-mail:] {\tt carlo.angelantonj@unito.it}
\\ {\tt ignatios.antoniadis@upmc.fr}
\\ {\tt marine.samsonyan@cern.ch}
\end{itemize}
}

\end{titlepage}

\setstretch{1.1}

\tableofcontents


\section{Introduction}

$\mathscr{N}=2$ supersymmetric theories, with two four-dimensional (4d) supercharges, provide a simple and interesting playground for studying exact  dynamics of gauge theories, their couplings to supergravity and various dualities. A very powerful tool is their relation with topological field and string theories, obtained by the so-called topological twist that combines R-symmetry with Lorentz transformations to define a projection into the chiral (BPS) sector of the theory~\cite{Witten:1988xj,Witten:1989ig}. 
In particular, $\mathscr{N}=2$ topological strings describe the coupling of topological field theories to gravity and its partition function computes a series of higher dimensional F-terms, $F_gW^{2g}$, involving powers of the chiral Weyl superfield $W$ in the effective $\mathscr{N}=2$ supergravity action~\cite{Antoniadis:1993ze, Bershadsky:1993cx}. Its lower component is the graviphoton field-strength (anti-self-dual by convention) that plays therefore the role of the topological string coupling. $F_0$ is the $\mathscr{N}=2$ prepotential, while $F_1$ corresponds to the gravitational $R^2$ coupling (on-shell there is only one term).

The field theory limit is obtained by going near the singularity of the moduli space where some charged states become massless depending on the gauge group and the representation matter content of the theory~\cite{Antoniadis:1995zn}. 
This limit is captured by the Nekrasov partition function, obtained by an explicit sum over instantons on the field theory side~\cite{Moore:1997dj,Losev:1997wp, Nekrasov:2002qd, Nekrasov:2003rj}. The sum is regularised by the so-called $\varOmega$-background, starting from six dimensions, and depends on two parameters $\epsilon_{1,2}$, or equivalently $\epsilon_\pm=\epsilon_1\pm\epsilon_2$. It was observed that for $\epsilon_+=0$, the power series in $\epsilon_-$ reproduces the (field theory limit of) $F_g$'s, leading to the identification of $\epsilon_-$ with the (anti-self-dual) graviphoton field-strength. $\epsilon_+$ provides therefore a deformation of the topological string, called refinement, and should be likely identified with a self-dual field strength of a particular vector modulus~\cite{Antoniadis:2013epe, Antoniadis:2013mna}.

A special case on the field theory side is that of an Abelian gauge theory which, despite the naive expectation, is actually non-trivial, even at the perturbative level, in the presence of a massive neutral hypermultiplet that makes it $\mathscr{N}=2^*$~\cite{Nekrasov:2003rj, Poghossian:2008ge}. The latter is defined as a massive deformation of  $\mathscr{N}=4$: when the mass $m$ of the hypermultiplet vanishes, one recovers $\mathscr{N}=4$ supersymmetry and a vanishing partition function, while when $m$ becomes very large, the adjoint hypermultiplet decouples, and one is left with a pure $\mathscr{N}=2$ $\textrm{U}(1)$ gauge theory. Despite the absence of renormalizable interactions and of singular points with extra massless states, the theory couples to gravity leading to non-trivial higher-dimensional couplings involving the graviphoton. Moreover $F_1$ has a logarithmic singularity in the mass, due to the non-vanishing trace anomaly. Finally, one can get another degree of non-triviality by considering a radius deformation from five dimensions.

In this work we present a string realisation of an Abelian 5d and 4d $\mathscr{N}=2^*$ theory~\footnote{For the non-Abelian case, see Ref.~\cite{Florakis:2015ied}.} and compute the couplings $F_{g,n}$ of a double series of higher-dimensional F-terms of the form $W^{2g}\Pi^{n}$, where $\Pi$ is the chiral projection of a certain anti-chiral vector superfield with lower component corresponding to the self-dual field strength. This amounts to compute a series of amplitudes involving four gravitini, $2g-2$ anti-self-dual graviphotons and $n$ self-dual gauge fields belonging to the multiplet of the D5-brane coupling modulus. In the field theory limit, the result reproduces the Nekrasov partition function in the two-parameter $\varOmega$-background, in agreement with the proposal~\cite{Antoniadis:2013epe}.

Our paper is organised as follows. In Section~\ref{freeactZN}, we present the string theory construction in a D-brane setup~\cite{Angelantonj:2002ct,Dudas:2000bn}, based on a freely acting orbifold that realises partial supersymmetry breaking $\mathcal{N}=4\to\mathcal{N}=2$~\cite{Antoniadis:1998ep}, as a Scherk-Schwarz deformation~\cite{Scherk:1979zr}. In Section~\ref{oneparameter}, we study the simplest case of $\epsilon_+=0$, corresponding to the usual topological amplitudes $F_g$ involving only powers of the graviphoton. We start by the 5d theory compactified on a circle of radius $R$ and then we take the 4d limit $R\to 0$. In Section~\ref{twoparameter}, we generalise our analysis to the deformed case described by the two-parameter $\varOmega$ background. In Section~\ref{nonabelian}, we extend the previous results to the case of non-Abelian $\textrm{U}(M)$ gauge theories thus recovering the results of \cite{Florakis:2015ied} and generalising them to the case of the two-parameter $\varOmega$ background. Finally, in Section~\ref{concls}, we present our conclusions and comment on the inclusion of non-perturbative contributions.

\section{The construction of $\mathscr{N}=2^*$ gauge theories} \label{freeactZN}

The $\mathscr{N}=2^*$ supersymmetric gauge theory is a massive deformation of  $\mathscr{N}=4$  whereby four scalars and two fermions, filling an $\mathcal{N}=2$ hypermultiplet, in the adjoint representation of the gauge group get a mass $m$. For generic values of the mass parameter, the hypermultiplet participates non-trivially to the dynamics of the gauge theory, while in the $m\to \infty$ limit the adjoint hypermultiplet decouples, and one is left with a pure $\mathscr{N}=2$ gauge theory. In the opposite $m\to 0 $ limit one simply recovers the original  $\mathscr{N}=4$ theory, where non-trivial cancellations occur and the theory becomes conformal at the quantum level. 

This property of the theory reminds of a spontaneous partial supersymmetry breaking via a Scherk-Schwarz mechanism \cite{Scherk:1979zr} where the mass of the adjoint hypermultiplet is proportional to the (inverse) size of the compact direction. In the following we shall build the $\mathcal{N}=2^*$ theory as a Scherk-Schwarz deformation of the $\mathcal{N}=4$ one, first in the field theory and then in a full-fledged string setup. 

\subsection{The Scherk-Schwarz construction in field theory}

It is a well known fact \cite{Gliozzi:1976qd} that the $\mathcal{N}=4$ gauge theory can be obtained as a dimensional reduction of the ten-dimensional $\mathcal{N}=(1,0)$ theory. Imposing periodic boundary conditions on all fields along the six compact directions, one is clearly left with a low-energy four-dimensional theory comprising one vector, six scalars and four Majorana fermions, thus filling the $\mathcal{N}=4$ vector multiplet. Following \cite{Scherk:1979zr} one can actually extend this set-up to allow for non-trivial periodicity conditions. The simplest construction compatible with eight supercharges,  amounts to splitting $T^6 = T^4 \times S^1 (R_5) \times S^1 (R_4)$ and impose on the various fields the boundary conditions
\begin{equation}
\phi (x_4+2 \pi R_4) = e^{i Q}\, \phi (x_4)\,,
\end{equation}
where $x_4$ is the compact coordinate along the Scherk-Schwarz circle $S^1 (R_4)$ and $Q$ is a symmetry charge that rotates the coordinates on the $T^4$ in a way that preserves supersymmetry. For instance, if we decompose $T^4 =T^2 \times T^2$ and $z_1$ and $z_2$ are the two complex coordinates on the two $T^2$'s, $Q$ acts as $z_1 \to e^{i\alpha} z_1$ and $z_2 \to e^{-i\alpha} z_2$. This action on the coordinates clearly induces a natural action on the components of the gauge fields along the $T^4$ and on their fermionic partners. 

As it is well known \cite{Scherk:1979zr}, this coordinate-dependent compactification induces shifts of the $p_4$ momenta $p_4 \to p_4 + \alpha /R_4$ and thus induces a mass gap for the hypermultiplet proportional to $\alpha/R_4$, while the Kaluza-Klein masses only scale like $1/R_4$. Since in the field theory, the boost of the Scherk-Schwarz deformation is arbitrary, one can properly decouple the two scales and a limit exists where all Kaluza-Klein states are decoupled while the massive hypermultiplet still participates to the dynamics. It is in this limit that one recovers the $\mathscr{N}=2^*$ gauge theory. 

Notice that in general, the Scherk-Schwarz deformation may act non-trivially on the gauge group degrees of freedom thus breaking  the original gauge group $G\to G_1 \times G_2$. In this case, one is left with an  $\mathcal{N}=2^*$ theory with gauge group $G_1 \times G_2$ coupled to a massless hypermultiplet in the bi-fundamental representation. To avoid these extra states we have chosen the Scherk-Schwarz deformation to have a trivial action on the gauge degrees of freedom. Moreover, in the following we shall restrict our attention to the Abelian case, deferring a discussion of the non-Abelian theory to the conclusions.

\subsection{The Scherk-Schwarz construction in string theory}\label{SSString}

An easy way to realise $\mathscr{N}=2^*$ in string theory is in terms of D-branes. In order to give mass to the hypermultiplet we employ the Scherk-Schwarz deformation discussed previously in the field theory context. Concretely, we use the well-known connection between Scherk-Schwarz reductions and freely-acting orbifolds \cite{Rohm:1983aq, Kounnas:1988ye, Kounnas:1989dk}, whereby the non-trivial boundary conditions are traded for a simultaneous action of the rotation on the $T^4$ coordinates and a shift along the $S^1 (x_4 )$ circle.

As before, in order to make contact with the Nekrasov computation of the perturbative free-energy on the $\varOmega$ background, we are actually interested in the field-theory limit of suitable string amplitudes. As a result, we need  to introduce a clear hierarchy within the Kaluza-Klein and Scherk-Schwarz masses that allows for a consistent decoupling limit. This is in principle a problem since, in string theory, continuous global symmetries do not exist and they are typically broken to discrete ones. Moreover, a compact $T^4$ together with the requirement of supersymmetry implies that the allowed symmetries are the discrete rotations $\mathbb{Z}_N$, with $N=2,3,4,6$.  As a result, in this setup , the Kaluza-Klein scale and the Scherk-Schwarz scale do not decouple \cite{Antoniadis:1998ep} and the field theory limit does not yield $\mathscr{N}=2^*$.

This problem could be circumvented if a $\mathbb{Z}_N$ with arbitrarily large $N$ were allowed. For this reason we shall assume that the $T^4$ be replaced by the non-compact space $\mathbb{C}^2$. This choice actually leads automatically to the decoupling of the gravitational sector, since the four-dimensional Newton constant vanishes and, as we shall see, yields the desired gauge theory in five and/or four dimensions. 

Taking this into account, our set-up consists then of a single D5-brane placed on the $\mathbb{C}^2$ origin of  the  $\mathscr{M}_{1,3}\times S^1_m \times S^1_R \times \mathbb{C}^2/\mathbb{Z}_N$ space.  Here, with a slight change of notation, $S^1_m$ is the Scherk-Schwarz circle with radius $R_{\rm SS} = m^{-1}$, while $S^1_R $ is a spectator circle. For generic values of its radius $R$ one describes a five-dimensional $\mathcal{N}=2^*$ gauge theory, while the four-dimensional theory is recovered in the  $R\to 0$ limit.

The orbifold group $\mathbb{Z}_N$ acts on the two complex coordinates $(z_1, z_2) \in \mathbb{C}^2$ as the rotation 
\begin{equation}
(z_1 \,,\, z_2 ) \to (e^{2 i \pi /N} z_1 \,,\, e^{-2i\pi/N} z_2 )\,.
\end{equation}
The fact that the twist is opposite in the two planes preserves $\mathscr{N}=2$ supersymmetries and, because the space is non-compact, there is no restriction on the order $N$. The mass of the hypermultiplet is then generated by combining the $\mathbb{Z}_N$ rotation with an order-$N$ shift along the Scherk-Schwarz circle $S^1_m$
\begin{equation}
S^1_m \ni y_m \to y_m + \frac{2\pi}{N m}\,.
\end{equation}

Note that the non-compactness of the space $\mathbb{C}^2 /\mathbb{Z}_N$ is also required for the consistency of this minimal D-brane construction. In fact, were the space compact one should have imposed tadpole conditions for the (untwisted) RR forms \cite{Angelantonj:2002ct,Dudas:2000bn}. Their cancellation would have called for an orientifold plane or for  anti-brane that would have compromised the simple $\mathcal{N}=2^*$ gauge theory construction. Twisted tadpole conditions, which should be imposed independently of the compactness or not of the transverse space, are also absent in this construction since, because of the free action of the Scherk-Schwarz deformation, only massive states with odd winding number do propagate in the transverse channel. 

This said, the partition function associated to a single D5 brane then reads
\begin{equation}
\mathscr{A} = \frac{1}{N} \left[ \sum_{\ell =0}^{N-1} \rho \big[ {\textstyle{0\atop \ell}}\big] \, \sum_{r\in\mathbb{Z}} e^{2 i \pi r\ell/N}\, P_r (1/m)\right] \sum_{s\in\mathbb{Z}} P_s (R)\,,
\label{annulusSSS}
\end{equation}
Here, $P_n (\rho ) = q^{\frac{1}{2} (n/\rho)^2} = e^{-\pi t (n/\rho)^2}$ is the generic contribution of the quantised momenta, while we have introduced the compact notation $ \rho \big[{\textstyle{0 \atop \ell}}\big]$ to encode the contribution of the world-sheet bosons and fermions, where
\begin{equation}
 \rho \big[{\textstyle{0 \atop 0}}\big] = \tfrac{1}{2}\, \sum_{a,b=0,1} (-1)^{a+b+ab}\, \frac{\theta^4\big[{a/2\atop b/2}\big]}{\eta^{12}}\,,
\end{equation}
and
\begin{equation}
 \rho \big[{\textstyle{0 \atop \ell}}\big] =\tfrac{1}{2}\, \sum_{a,b=0,1} (-1)^{a+b+ab}\, \frac{\theta^2 \big[{a/2\atop b/2}\big]}{\eta^{6}}\, \left( 2 \sin (\pi\ell/N )\, \frac{\theta \big[{a/2\atop b/2+\ell/N}\big]}{\theta \big[{1/2\atop 1/2+\ell/N}\big]}\right)\,  \left( 2 \sin (-\pi\ell/N )\, \frac{\theta \big[{a/2\atop b/2-\ell/N}\big]}{\theta \big[{1/2\atop 1/2-\ell/N}\big]}\right) \,,
\end{equation}
for $\ell\not=0$.

To analyse the spectrum it is convenient to separate
\begin{equation}
 \rho \big[{\textstyle{0 \atop \ell}}\big] = \beta \big[{\textstyle{0 \atop \ell}}\big] - \varphi \big[{\textstyle{0 \atop \ell}}\big]\,,
\end{equation}
into space-time bosons, the $\beta \big[{\textstyle{0 \atop \ell}}\big]$ with $a=0$, and space-time fermions, the $\varphi \big[{\textstyle{0 \atop \ell}}\big]$ with $a=1$. Moreover, it suffices to concentrate on the bosons since the fermions will follow by supersymmetry. We write
\begin{equation}
\beta \big[{\textstyle{0 \atop 0}}\big] = V_4\, \frac{O_4}{\eta^{10}} + O_4 \, \frac{V_4}{\eta^{10}}\,,
\end{equation}
and
\begin{equation}
\beta \big[{\textstyle{0 \atop \ell}}\big] = V_4\, \frac{O_4 (\ell/N)}{\eta^4} + O_4 \, \frac{V_4 (\ell /N)}{\eta^4}\,,
\label{beta0l}
\end{equation}
where
\begin{equation}
\begin{split}
O_4 (\ell/N) &= \tfrac{1}{2} \left[ \left(2 \sin (\pi \ell /N) \frac{\vartheta_3 (\ell/N|\tau)}{\vartheta_1 (\ell/N|\tau)} \right)^2 + \left(2 \sin ( \pi \ell /N) \frac{\vartheta_4 ( \ell/N|\tau)}{\vartheta_1 ( \ell/N|\tau)} \right)^2  \right]\,,
\\
V_4 (\ell/N) &= \tfrac{1}{2} \left[ \left(2 \sin (\pi \ell /N) \frac{\vartheta_3 (\ell/N|\tau)}{\vartheta_1 (\ell/N|\tau)} \right)^2 - \left(2 \sin ( \pi \ell /N) \frac{\vartheta_4 ( \ell/N|\tau)}{\vartheta_1 ( \ell/N|\tau)} \right)^2  \right]\,,
\end{split}
\end{equation}
are the SO(4) level-one characters with non-trivial argument. A $q$-Taylor expansion of the previous expressions yields
\begin{equation}
\begin{split}
O_4 (\ell/N) &\simeq 1+ 2 \left[5 + 2 \cos (2\pi \ell /N) + \cos (4 \pi \ell /N))\, q +\ldots \right]\,,
\\
V_4 (\ell /N) &\simeq 4 \cos (2 \pi \ell /N ) + 8 \left[1+4 \cos (2\pi \ell /N) + \cos (4 \pi \ell /N)\right]\, q +\ldots \,.
\end{split}\label{chol}
\end{equation}
Combining this expansion for the characters of the internal SO(4) with their space-time counterpart, and taking into account that fermions have similar expansions as a consequence of supersymmetry, one arrives at the following expansion for the light states of the annulus partition function
\begin{equation}
\mathscr{A}_0 \simeq (V_4 - 2 S_4 )\, \left[ \frac{1}{N} \sum_{\ell =0}^{N-1} e^{2 i \pi r \ell/N} q^{\frac{1}{4} (rm)^2} \right]+ (4\, O_4 -2 C_4) \, \left[\frac{1}{N} \sum_{\ell=0}^{N-1} \cos (2\pi \ell/N ) \, e^{2 i \pi r\ell/N}\, q^{\frac{1}{4} (rm)^2}\right] \,.
\end{equation}
Since,
\begin{equation}
\frac{1}{N} \sum_{\ell =0}^{N-1} e^{2 i \pi r \ell/N} = \begin{cases} 1 & {\rm if}\ r=0\ {\rm mod}\ N\,,\\ 0 & {\rm otherwise}\,,\end{cases}
\end{equation}
and
\begin{equation}
\begin{split}
\frac{1}{N} \sum_{\ell=0}^{N-1} \cos (2\pi \ell/N ) \, e^{2 i \pi r\ell/N} &= \frac{1}{2N}\sum_{\ell=0}^{N-1} \left( e^{2 i \pi (r+1)\ell /N} + e^{2 i \pi (r-1)\ell /N} \right) 
\\
&= \begin{cases} \frac{1}{2} & {\rm if}\ r=\pm 1\ {\rm mod}\ N\,, \\ 0 & {\rm otherwise}\,,\end{cases}
\end{split}
\end{equation}
one finds that the vector multiplet has KK masses along the shifted direction given by
\begin{equation}
M^2_V = \tfrac{1}{2} (N k m)^2\,, \qquad k=0 , \pm 1 , \pm 2, \ldots
\end{equation}
whereas there are two hypermultiplets with masses
\begin{equation}
M^2_{H_1} = \tfrac{1}{2} (1+k N)^2 m^2\,, \qquad k=0,1,2,\ldots\,,
\end{equation}
and
\begin{equation}
M^2_{H_2} = \tfrac{1}{2} (1-kN)^2 m^2\,, \qquad k=1,2,\ldots\,.
\end{equation}
If we take $N$ very large the second set of hypermultiplets becomes very massive, and we are left with an Abelian massless vector multiplet and a neutral hypermultiplet with mass $\tfrac{1}{2} m^2$.

\section{The one parameter deformation}\label{oneparameter}

In the previous Section we have constructed the Abelian $\mathcal{N}=2^*$ theory and derived its partition function in the flat Minkowski space. Following Nekrasov \cite{Nekrasov:2002qd} the gauge theory on the non-trivial $\varOmega$ background amounts at computing a set of topological amplitudes \cite{Antoniadis:1993ze,Antoniadis:1995zn} involving two gravitons and a certain number of (anti-)self-dual graviphotons or, alternatively, the topological string partition function \cite{Bershadsky:1993cx, Witten:1988xj,Witten:1989ig}. We start in this Section with the simple $\varOmega$ background depending on a single parameter $\hbar = \epsilon_1 = - \epsilon_2$.

\subsection{The topological amplitude}  \label{topampl}

Let us compute the topological amplitudes associated to the $R^2 F^{2g-2}$ higher-derivative coupling in the vacuum described in Section \ref{freeactZN}. Here $R$ is the self-dual Riemann tensor and $F$ is the anti-self-dual  graviphoton field-strength, where space-time indices are omitted for notational simplicity. Since we are interested in the behaviour of the gauge sector, it suffices to compute the previous amplitude on the Riemann surface with the topology of an annulus. Moreover, it is simpler to compute the supersymmetry-related amplitude involving four gravitini and $2g-2$ anti-self-dual graviphotons. 
\begin{equation}
\mathscr{A}_g = \left\langle (V_{\rm grav}^+)^2 \, (V_{\rm grav}^-)^2 \, V_{\rm gph}^{2g-2} \right\rangle \,.
\label{ampli}
\end{equation}

The vertex operators for the gravitini and the graviphotons are given by
\begin{eqnarray}
V_{\rm grav} ^{\pm} (\xi_{\mu\alpha} , p) &=& \xi_{\mu\alpha} e^{-\varphi/2} S^\alpha e^{i\phi_3/2} \sigma^{\pm} \, \left( \bar\partial Z^\mu + i (p\cdot \tilde\chi )\tilde\chi^\mu \right)\, e^{i p \cdot Z}\,, \label{gravitiniV}
\\
V_{\rm gph} (\epsilon , p) &=& \epsilon_\mu \Big[ (\partial X + i (p\cdot \chi) \psi ) (\bar\partial Z^\mu + i (p\cdot \tilde\chi )\tilde\chi^\mu )  \nonumber
\\
& &\qquad - e^{- (\varphi + \tilde\varphi )/2} p_\nu S^\alpha (\sigma^{\mu\nu} )_\alpha{}^\beta \tilde S_\beta \, e^{i (\phi_2 + \tilde \phi_3 )/2} \, \Sigma^+ \tilde \Sigma^- \Big]\, e^{i p \cdot Z} + ({\rm left} \leftrightarrow {\rm right}) \,.
\label{vertexops}
\end{eqnarray}
Here, $Z^{1,2}, X, Z^{4,5}$ denote the complexified bosonic coordinates of the non-compact space-time, the $T^2$ and the non-compact $\mathbb{C}^2$ directions transverse to the branes which are acted upon by the $\mathbb{Z}_N$ twist, respectively. Similarly for the fermionic fields $\chi^{1,2} , \psi, \chi^{4,5}$, and  $S , e^{\pm i \phi_3/2}, \Sigma$ denote the corresponding spin fields in the NS (Neveu-Schwarz) and R (Ramond)sectors, respectively, while the tilde refers to the right-moving sector of the closed string (arbitrarily defined). Finally, $\varphi$ is associated to the bosonised superghost. 

Following \cite{Antoniadis:1993ze,Antoniadis:1995zn}, it is convenient to choose the polarisations and the momenta of the $2g+2$ vertex operators as follows
\begin{equation}
V_{\rm grav}^+ (\xi_{21}, p_1)\,, \qquad V_{\rm grav}^+ (\xi_{22} , p_{\bar 1} )\,, \qquad V_{\rm grav}^- (\xi_{\bar 2 1}, p_1)\,,  \qquad V_{\rm grav}^- (\xi_{\bar 2 2}, p_{\bar 1} )
\end{equation}
for the four gravitini, and
\begin{equation}
V_{\rm gph} (\epsilon_2 , p_1) \,, \qquad V_{\rm gph} (\epsilon_{\bar 2} , p_{\bar 1})
\end{equation}
for the $2g-2$ graviphotons. 

With these kinematic factors, upon performing the sum over the various spin structures, the amplitude \eqref{ampli} reduces to correlators in the odd spin structure involving only  fields
with indices along the space-times directions. These do not depend on the orbifold twist acting on the internal coordinates, and thus the corresponding correlators can be extracted from the generating functions
\begin{equation}
G_{\rm Bose} (\hbar ) = \left\langle \exp \frac{\hat\hbar}{t} \int d^2 \sigma \left[ Z^1 (\bar\partial - \partial )Z^2 + \bar Z^1 ( \bar\partial - \partial ) \bar Z^2 \right] \right\rangle 
= \frac{(\pi \hbar )^2}{\sin^2 (\pi \hat\hbar )} \left[ H_1 (\hat\hbar ; t/2) \right]^{-2} \,,
\end{equation}
for the bosonic coordinates, and
\begin{equation}
G_{\rm Fermi} (\hbar ) =\left\langle \exp \frac{\hat\hbar}{t} \int d^2 \sigma \left[ (\chi^1-\tilde\chi^1 )(\chi^2 - \tilde \chi^2 ) +(\bar\chi^1-\tilde{\bar \chi}^1 )(\bar \chi^2 - \tilde{\bar \chi}^2 ) \right]  \right\rangle = \left[ H_1 (\hat\hbar ; t/2) \right]^{2}\,,
\end{equation}
for the fermionic fields. Here we have introduced the dressed deformation $\hat\hbar = 2\hbar t\, (r m -i s/R)$, and
\begin{equation}
H_1 (z ; \tau ) = \frac{\theta_1 (z |\tau )}{2 \sin (\pi z)\, \eta^3 (\tau )} \prod_{m\in\mathbb{Z}\atop n>0} \left(1 - \frac{z^2}{|m+ z - n \tau |^2}\right)\,.
\end{equation}

In the odd spin structure, the contribution of the twisted world-sheet fermions cancels against the contribution of the twisted world-sheet bosons, and thus the generating amplitude simply reads
\begin{equation}
\mathscr{F} (\hbar ) = - \frac{4}{N} \sum_{\ell =1}^{N-1} \sum_{r,s\in\mathbb{Z}} \sin^2 \left(\frac{\pi \ell}{N}\right) \, e^{2 i \pi r\ell /N}\,  \int_0^\infty \frac{dt}{t} \, \frac{(\pi \hbar )^2}{\sin^2 (\pi \hat\hbar )}\,   P_r (1/m)\, P_s (R)\,.
\end{equation}
Using simple trigonometry,  we can cast the previous expression in the form
\begin{equation}
\begin{split}
\mathscr{F} (\hbar ) &= - 2 \sum_{r,s\in\mathbb{Z}} \left[ \frac{1}{N} \sum_{\ell =1}^{N-1}  \left( 1- \cos \,(2 \pi \ell/N ) \right) \, e^{2 i \pi r\ell /N} \right]\,  \int_0^\infty \frac{dt}{t} \, \frac{(\pi \hbar )^2}{\sin^2 (\pi \hat\hbar )}\,   P_r (1/m)\, P_s (R)
\\
&= - 2 \sum_{r,s\in\mathbb{Z}} \left[ \frac{1}{N} \sum_{\ell =0}^{N-1}  \left( 1- \cos \,(2 \pi \ell/N ) \right) \, e^{2 i \pi r\ell /N} \right]\,  \int_0^\infty \frac{dt}{t} \, \frac{(\pi \hbar )^2}{\sin^2 (\pi \hat\hbar )}\,   P_r (1/m)\, P_s (R)
\\
&= \left[ -2 \sum_{r=0\, {\rm mod} \, N\atop s\in\mathbb{Z}} + \sum_{r=\pm 1 \, {\rm mod}\, N\atop s\in\mathbb{Z}} \right]\,  \int_0^\infty \frac{dt}{t} \, \frac{(\pi \hbar )^2}{\sin^2 (\pi \hat\hbar )}\,   P_r (1/m)\, P_s (R) \,,
\end{split}
\end{equation}
where in the second step we have extended at no cost the sum over $\ell$ to include the $\ell =0$ term, while in the last step we have explicitly enforced the 
constraint on the KK momenta along the Scherk-Schwarz direction. The string excitations do not participate in this BPS saturated amplitude, and the field theory limit corresponds just to the $N\to \infty$ limit, where the Kaluza-Klein excitations along the Scherk-Schwarz direction decouple, leaving only a massless vector multiplet and a neutral hypermultiplet with mass $m$. They correspond, respectively,  to $r=0$ and $r=\pm1$  in
\begin{equation}
\begin{split}
\mathscr{F} (\hbar ) &=  \, (\pi\hbar )^2 \, (-2\delta_{r,0} + \delta_{r,1} +\delta_{r,-1}) \int_0^\infty \frac{dt}{t} \,  P_r (1/m)\, \sum_{s\in\mathbb{Z}} \frac{ P_s (R)}{\sin^2 
(\pi \hat\hbar )}
\\
&\equiv \, (\pi\hbar )^2 \, \sum_{\mu = 0 , \pm m} d(\mu )\,  F (\hbar , \mu ) \,,
\end{split}\label{integral}
\end{equation}
with  $d (0) = - 2$ and $d (\pm m ) =1$.

\subsection{Evaluating the integral} \label{evint}

Although the full amplitude $\mathscr{F} (\hbar )$ is finite, each individual contribution $F (\hbar , \mu)$ is not. The integrals \eqref{integral} are in fact divergent in the UV $(t\to 0)$ and thus need a proper regularisation. To this end, it is convenient to perform the change of variable
\begin{equation}
\begin{split}
\frac{\pi t}{R^2}  (s+i\mu R) \to t 
\qquad & {\rm for}\quad s\ge 0\,,
\\
-\frac{\pi t}{R^2}  (s+i\mu R ) \to t
\qquad & {\rm for}\quad s< 0\,,
\end{split}\label{changevar}
\end{equation}
so that eq. (\ref{integral}) becomes
\begin{equation}
F (\hbar , \mu ) = - \int_0^\infty \frac{dt}{t} \frac{1}{\sinh^2 (2 R \hbar t)} \left[ \sum_{s=0}^\infty e^{-t (s-i\mu R)}+ \sum_{s=1}^\infty e^{-t (s+i\mu R)} \right].
\end{equation}

The sums over $s$ can be easily performed, while it is convenient to expand
\begin{equation}
\frac{1}{\sinh^2 (2 R \hbar t )} =  - 4 \sum_{g=0}^\infty\frac{B_{2g} \, (2g-1)}{(2g)!}\, (4 R\hbar t )^{2g-2}\,, 
\end{equation}
with $B_{2g}$ the Bernoulli numbers. As a result,
\begin{equation}
F (\hbar , \mu ) =   \sum_{g=0}^\infty F_{2g} (\hbar , \mu) = 4 \sum_{g=0}^\infty  \frac{B_{2g} \, (2g-1)}{(2g)!}\, (4 R\hbar )^{2g-2} \, \int_0^\infty \frac{dt}{t}  \, t^{2g-2}\,  \frac{e^{i t \mu R } + e^{-t (1+i\mu R)}}{1-e^{-t}} \,.
\end{equation}
The generic integral 
\begin{equation}
I_g = \int_0^\infty \frac{dt}{t} \, t^{2g-2} \, \frac{e^{-at}}{1-e^{-t}}\,,
\end{equation}
is convergent for $g>1$, and admits a well known analytical continuation to arbitrary $g$ on the complex plane by deforming the integration domain to the Hankel contour
\begin{equation}
\tilde I _g = -\frac{1}{2 i \sin (\pi (2g-2))} \int_\infty^{(0^+)} dt\, (-t)^{2g-3} \, \, \frac{e^{-at}}{1-e^{-t}} \,.
\end{equation}
Taking into account the definition of the Hurwitz zeta function,
\begin{equation}
\zeta (\sigma ;a ) = - \frac{\varGamma (1-\sigma )}{2\pi i} \int_\infty^{(0^+)} dt \, (-t)^{\sigma-1}\, \frac{e^{-at}}{1-e^{-t}}\,,
\end{equation}
the generic term in the sum is thus
\begin{equation}
F_{2g} ( \hbar , \mu ) = 4\,  \frac{B_{2g} \, (2g-1)}{(2g)! }\, (4 R \hbar )^{2g-2} \, \varGamma (2g-2)\, \left[ \zeta (2g-2 ; - i \mu R ) + \zeta (2g-2 ; 1 + i \mu R ) \right]\,.
\end{equation}
This combination is regular for $g>1$, while care is needed to extract the $g=0$ and $g=1$ contributions. For the latter one has
\begin{equation}
\begin{split}
& 4 \frac{(2g-1)}{\varGamma (2g+1)} (4 R \hbar )^{2g-2}\, \Gamma (2g-2 ) \left( \zeta (2g-2; - i \mu R) + \zeta (2g-2 ; 1 + i \mu R )\right)
\\
& \qquad \sim \frac{1}{g-1} \left( \zeta (0;-i\mu R) + \zeta (0;1+i\mu R )\right) - \left( 1 - 2 \log (4 R \hbar ) \right)\left( \zeta (0;-i\mu R ) + \zeta (0;1+i\mu R)\right) 
\\
&\qquad+ 2 \left( \zeta ' (0;-i\mu R) + \zeta ' (0;1+i \mu R )\right)\,,
\end{split}
\end{equation}
as $g\to 1$, and
\begin{equation}
\begin{split}
4 \frac{(2g-1)}{\varGamma (2g+1)}& (4 R \hbar )^{2g-2}\, \Gamma (2g-2 ) \left( \zeta (2g-2;- i\mu R) + \zeta (2g-2 ; 1 + i \mu R )\right)
\\
& \qquad \sim -\frac{1}{(4 R\hbar )^2}\, \Bigg[  \frac{1}{g } \left( \zeta (-2;-i\mu R) + \zeta (-2;1+i\mu R )\right) 
\\
& \qquad+ \left( 3 -2 \gamma + 2 \log (4 R \hbar ) \right)\left( \zeta (-2;-i\mu R) + \zeta (-2;1+i\mu R)\right) 
\\
&\qquad+ 2 \left( \zeta ' (-2;-i\mu R) + \zeta ' (-2;1+i \mu R )\right)\Biggr]\,,
\end{split}
\end{equation}
as $g\to 0$. 
The Hurwitz zeta function satisfies remarkable identities, among which
\begin{equation}
\begin{split}
\zeta (-m;a) &= - \frac{B_{m+1} (a)}{m+1}\,, 
\\
\zeta ' (0;a) &= \log \varGamma (a) - \tfrac{1}{2}\log (2\pi )\,, 
\end{split}
\qquad
\begin{split}
\zeta (0;a ) &= \tfrac{1}{2} - a \,, 
\\
\zeta (-2 ; - ia ) + \zeta (-2;1+ia) &=0\,,
\end{split}
\end{equation}
where $B_n (a)$ is the Bernoulli polynomial. Moreover, 
\begin{equation}
\begin{split}
- \frac{2}{(4 \hbar R)^2} \left[ \zeta ' (-2 ; -i \mu R ) + \zeta ' (-2 ; 1 +i \mu R )\right] &=  \frac{1}{2(4 \pi \hbar R)^2} \left[ {\rm Li}_3 \left(e^{2\pi \mu R } \right) + {\rm Li}_3 \left( e^{-2\pi \mu R}\right)\right]
\\
&=  \frac{1}{ (4 \pi \hbar R)^2} \left[ {\rm Li}_3 \left( e^{-2\pi \mu R}\right) + \tfrac{2}{3} i \pi^3 \, B_3 (- i \mu R)\right]
\\
&= \frac{1}{(4 \pi \hbar R)^2} \left[  {\rm Li}_3 \left( e^{-2\pi \mu R}\right) - \tfrac{1}{12} (2 \pi \mu R )^3 \right.
\\
&\qquad \qquad \left. + \tfrac{1}{4} i \pi ( 2 \pi \mu R )^2 +\tfrac{1}{6} \pi^2\, ( 2\pi \mu R ) \right] \,.
\end{split}
\end{equation}
Combining all this, one finally gets
\begin{equation}
F_2 (\hbar , \mu) = - \tfrac{1}{3} \log \left( 2 \, \sinh \left( \pi \mu R \right)\right) + \frac{i\pi}{6} \,,
\end{equation}
and
\begin{equation}
F_0 (\hbar , \mu) = \frac{1}{(4 \pi \hbar R)^2} \left[  {\rm Li}_3 \left( e^{-2\pi \mu R}\right) - \tfrac{1}{12} (2 \pi \mu R )^3 
+ \tfrac{1}{4} i \pi ( 2 \pi \mu R )^2 +\tfrac{1}{6} \pi^2\, ( 2\pi \mu R ) \right] \,.
\end{equation}

As a result, one finds 
\begin{equation}
\begin{split}
F (\hbar , \mu ) &= \frac{1}{(4 \pi \hbar R)^2} \left[  {\rm Li}_3 \left( e^{-2\pi \mu R}\right) - \tfrac{1}{12} (2 \pi \mu R )^3 + \tfrac{1}{4} i \pi ( 2 \pi \mu R )^2 +\tfrac{1}{6} \pi^2\, ( 2\pi \mu R ) \right]
\\
&-\tfrac{1}{3} \log \left( 2 \, \sinh \left( \pi \mu R \right)\right) + \frac{i\pi}{6}
\\
&+4\sum_{g=2}^\infty \frac{B_{2g}}{2g \, (2g-2)} (4 R \hbar)^{2g-2} \, \left( \zeta (2g-2 ; - i \mu R ) + \zeta (2g-2 ; 1 + i \mu R )\right) \,.
\end{split}\label{FhbarHur}
\end{equation}
Different expressions are possible for $F (\hbar , \mu )$. For instance,  noting that
\begin{equation}
 \zeta (2g-2 ; - i m Rr ) + \zeta (2g-2 ; 1 + i m R r ) = \sum_{n\in\mathbb{Z}} \frac{1}{(n-imRr )^{2g-2}}
= \frac{(2\pi i )^{2g-2}}{(2g-3)!} \, {\rm Li}_{3-2g} \left( e^{-2\pi mRr}\right)\,,
 \end{equation}
 where in the last equality we have Poisson summed the series, one can recast eq. \eqref{FhbarHur} as
 \begin{equation}
 \begin{split}
F (\hbar , \mu)  &=  \frac{1}{(4 \pi \hbar R)^2} \left[  {\rm Li}_3 \left( e^{-2\pi \mu R}\right) - \tfrac{1}{12} (2 \pi \mu R )^3 + \tfrac{1}{4} i \pi ( 2 \pi \mu R )^2 +\tfrac{1}{6} \pi^2\, ( 2\pi \mu R ) \right]
\\
&- \tfrac{1}{3} \log \left( 2 \, \sinh (\pi \mu R  )\right) + \frac{i\pi}{6}
+ 4\sum_{g=2}^\infty \frac{B_{2g}}{2g \, (2g-2)!} (8\pi i R \hbar)^{2g-2} \, {\rm Li}_{3-2g} \left(e^{-2\pi \mu R} \right) \,. 
\end{split} \label{freeen}
\end{equation}

The contribution of the massive hypermultiplet, corresponding to $\mu=\pm m $, can be straightforwardly derived from the previous expression. The contribution of the vector multiplet is a bit subtler since setting $\mu =0$ would yield a logarithmic divergence in the $g=1$ term. This is precisely what one would expect from a field theory viewpoint and a suitable regularisation yields\footnote{In the following expression we have used the fact that, for the vector multiplet contribution, the only dimensionless combination in the constant term in the $\hbar$ expansion is $\varLambda R$.}
\begin{equation}
- \tfrac{1}{3} \frac{d}{ds} \left[\frac{(\varLambda R)^s}{\Gamma (s)} \int_0^\infty \frac{dt}{t} t^s \, e^{-t} \right]_{s=0} = - \tfrac{1}{3} \log\,(\varLambda R)\,. \label{regular}
\end{equation}

\subsection{The perturbative free energy}

According to Nekrasov \cite{Nekrasov:2002qd}, the topological amplitude we have just considered should compute the perturbative contribution to the pre-potential of the $\mathcal{N}=2^*$ theory on the one-parameter $\varOmega$ background,
\begin{equation}
(\pi\hbar )^{-2}\, \mathscr{F} (\hbar ) = 8\, \left( \gamma_\hbar (m;R) - \gamma_\hbar (0;R ) \right)\,,
\end{equation}
with 
\begin{equation}
\gamma_\hbar (x;\beta ) = \frac{\beta x^3}{12 \hbar^2} - \frac{x^2}{2\hbar^2}\, \log (\beta \varLambda ) - \frac{\beta x}{24} + \sum_{n=1}^\infty \frac{1}{n}\, \frac{e^{-\beta n x}}{(e^{-\beta n \hbar }-1 )(e^{\beta n \hbar } -1)}\,.
\end{equation}
Indeed, upon the redefinitions 
\begin{equation}
2 \pi R \to R\,, \qquad 4 i \hbar \to \hbar\,,
\end{equation}
in eq. \eqref{freeen}, and the use of the identity 
\begin{equation}
\textrm{Li}_s (z) + (-1)^s\, \textrm{Li}_s (1/z) = \frac{ (2\pi i )^s}{\varGamma (s)}\, \zeta \left( 1-s ; \tfrac{1}{2} + \frac{\log (-z)}{2\pi i}\right)\,,
\end{equation}
one gets
\begin{equation}
\begin{split}
(\pi \hbar )^{-2}\, \mathscr{F} (\hbar ) &= \sum_{\mu = 0 ,\, \pm m} d (\mu ) \, F (\hbar , \mu )
\\
&= \frac{8}{\hbar^2 R^2} \left[ \zeta (3) - {\rm Li}_3 \left( e^{- m R}\right) + \tfrac{1}{12} (m R)^3 - \tfrac{1}{2} i \pi (mR)^2 - \tfrac{1}{6}\pi^2 mR \right]
\\
&+ \tfrac{2}{3} \left[ \log (\varLambda R) - \log \left( 2 \sinh \left(\frac{m R}{2}\right)\right) \right]
\\
&+8  \sum_{g=2}^\infty \frac{B_{2g}}{2g \, (2g-2)!} (R\hbar )^{2g-2}\,  \left[ {\rm Li}_{3-2g} \left( e^{-mR}\right) - \zeta (3-2g )\right]\,,
\end{split}\label{5dfreeen}
\end{equation}
that matches precisely the result of \cite{Nekrasov:2002qd, Nekrasov:2003rj}, up to irrelevant constants depending on the regularisation scheme.

\subsection{The four-dimensional theory}

As anticipated in Section \ref{freeactZN} the four-dimensional theory can be obtained as the $R\to 0$ limit, whereby the tower of Kaluza-Klein states along the spectator circle decouples and one is left with an effective four-dimensional model. In order to compute this limit, we note that
\begin{equation}
{\rm Li}_{-n} (z) = \frac{\partial^{n+1}}{\partial z^{n+1}} {\rm Li}_1 (z)
= - \frac{\partial^{n+1}}{\partial z^{n+1}}\, \log (1-z) 
=\frac{1}{(1-z)^{n+1}} \sum_{k=0}^{n-1} \left\langle {n\atop k}\right\rangle z^{n-k}
\end{equation}
where 
\begin{equation}
\left\langle {n\atop k}\right\rangle = \sum_{k=0}^{m+1} (-1)^k \left( {n+1 \atop k} \right) \, (m+1-k)^n
\end{equation}
are the Eulerian numbers. By taking $z=e^{-y}$, in the limit $y\to 0$ it is immediate to see that 
\begin{equation}
{\rm Li}_{-n} \left(e^{-y} \right) = \frac{1}{y^{n+1}} \sum_{k=0}^{n-1} \left\langle {n\atop k}\right\rangle + O (y^{-n}) = \frac{n!}{y^{n+1}} + O (y^{-n})\,, \label{limitpolylog}
\end{equation}
where we have used $\sum_{m=0}^{n-1} \langle {n\atop m}\rangle =n!$.  Inserting this behaviour in eq. \eqref{5dfreeen} we easily get
\begin{equation}
(\pi \hbar )^{-2}\, \mathscr{F} \to  \tfrac{2}{3} \log \left( \frac{\varLambda}{m} \right) +8 \sum_{g=2}^\infty \frac{B_{2g}}{2g \, (2g-2)} \, \left( \frac{\hbar}{m}\right)^{2g-2} \,, \label{4doneparfree}
\end{equation}
which indeed corresponds to the perturbative contribution of the Nekrasov partition function on the one-parameter $\varOmega$ background.

\section{The two parameter deformation}\label{twoparameter}

We now turn to the study of $\mathscr{N}=2^*$ theories in the general $\varOmega$-background with the two parameters switched on, and in particular to the topological amplitude associated to it\footnote{Backgrounds involving more general deformations have been discussed in \cite{Orlando:2011nc,Hellerman:2012zf,Lambert:2013lxa}.}. The two parameters correspond to rotations in the two SU(2)'s of the four-dimensional Lorentz group (properly combined with rotations on the SU(2) R-symmetry) and thus it is natural to associate them to self-dual and anti-self-dual configurations of gauge fields. Therefore, if the $\epsilon_-$ deformation is related to the insertion of an anti-self-dual graviphoton field strength in the topological amplitude, the second parameter $\epsilon_+$ can be introduced by adding insertions of self-dual configurations. Indeed, after many attempts \cite{Antoniadis:2010iq, Nakayama:2011be}, it was showed in \cite{Antoniadis:2013epe} that the correct topological amplitude now involves anti-self-dual graviphotons and self-dual backgrounds for the partners of gauge coupling on the D5 branes. 

\subsection{The topological amplitude}

Following \cite{Antoniadis:2013epe}, we now turn to compute the amplitude
\begin{equation}
\mathscr{A}_{g,n} = \left\langle (V_{\rm grav}^+)^2 \, (V_{\rm grav}^-)^2 \, V_{\rm gph}^{2g-2} V_{S'_+}^{2n}\right\rangle \,,
\label{ampli2parameters}
\end{equation}
where, as usual, it is more convenient to replace the two gravitons by four gravitini with vertex operator given in eq. \eqref{gravitiniV}. The vertex operator $V_{\rm gph}$ for the anti-self-dual graviphotons is again given by eq. \eqref{vertexops},  while 
\begin{equation}
\begin{split}
V_{S'_+} (\epsilon , p) &=\epsilon_\mu \, \Big[ (\partial X + i (p \cdot \chi)\,  \psi) \, (\bar{\partial} Z^\mu + i (p \cdot \tilde{\chi})\,  \tilde{\chi}^\mu)  
\\
&+  e^{-\frac{1}{2} (\varphi + \tilde{\varphi})} \, p_\nu \, S_{\dot{\alpha}} \, \left( \bar{\sigma}^{\mu \nu}\right) ^{\dot{\alpha}}_{\,\,\,\,\, \dot{\beta}}\, \tilde{S}^{\dot{\beta}} \, e^{\frac{i}{2}(\phi_3+\tilde{\phi}_3)} \, \hat{\Sigma} ^+ \, \hat{\tilde{\Sigma}}^-\Big] \, e^{i p \cdot Z} +({\rm left} \leftrightarrow {\rm right})\,,
\end{split} 
\end{equation}
is the vertex operator of the self-dual vector multiplet, whose scalar is the (complex conjugate) of the $S'$ modulus. 

As before, since we are interested in the one-loop amplitude with open-string fields running in the loop, it suffices to restrict our attention to the Riemann surface with the topology of an annulus. This amplitude is quite involved even when polarisation vectors and momenta are suitably chosen as in Section \ref{topampl}. This is a consequence of the fact that now all vertices contribute in all possible ways to the amplitude. Following \cite{Antoniadis:2013epe},  after tedious computations and careful regularisation of the one-loop determinants, one is lead to the generating functions
\begin{equation}
\begin{split}
G_{\textrm{ Bose}} (\epsilon_\pm) &= \left\langle \exp \left[\frac{\hat{\epsilon}_-}{t} \int d^2 \sigma \left[ Z^1 (\bar\partial - \partial )Z^2 + \bar Z^1 ( \bar\partial - \partial ) \bar Z^2 \right] 
+\frac{\hat{\epsilon}_+}{t} \int d^2 \sigma \left[ Z^1 (\bar\partial - \partial )\bar Z^2 + \bar Z^1 ( \bar\partial - \partial )  Z^2 \right]\right]  \right\rangle
\\
&=\left[H_1\left(\frac{\hat{\epsilon}_- -\hat{\epsilon}_+}{2}; 0 ; \frac{t}{2} \right) \, H_1\left(\frac{\hat{\epsilon}_- +\hat{\epsilon}_+}{2}; 0 ; \frac{t}{2} \right) \right]^{-1} \frac{\pi (\epsilon_- - \epsilon_+)}{\sin \pi (\hat{\epsilon}_- - \hat{\epsilon}_+)} \,  \frac{\pi (\epsilon_- + \epsilon_+)}{\sin \pi (\hat{\epsilon}_- + \hat{\epsilon}_+)}  , \label{Gbose}
\end{split}
\end{equation}
for the bosonic coordinates, 
\begin{equation}
G_{\textrm{Fermi}} (\epsilon_- ) =\left\langle \exp \frac{\hat{\epsilon}_-}{t} \int d^2 \sigma \left[ (\chi^1- \tilde\chi^1 )(\chi^2 -  \tilde \chi^2 ) +(\bar\chi^1- \tilde{\bar \chi}^1 )(\bar \chi^2 -\tilde{\bar \chi}^2 ) \right]  \right\rangle = \left[ H_1 \left(\frac{\hat{\epsilon}_-}{2} ;0;  \frac{t}{2}\right) \right]^{2}\,, \label{Gfermi}
\end{equation}
for the space-time fermions, and
\begin{equation}
\begin{split}
G_{\textrm{K3\ Fermi}} (\epsilon_+ ) &=\left\langle \exp \frac{\hat{\epsilon}_+}{t} \int d^2 \sigma \left[ (\chi^4 +\tilde\chi^4 )(\chi^5 + \tilde \chi^5 ) +(\bar\chi^4+ \tilde{\bar \chi}^4 )(\bar \chi^5 + \tilde{\bar \chi}^5 ) \right]  \right\rangle
\\
&=-4 \sin^2 \left(\frac{\pi \ell}{N}\right) \, H_1\left(\frac{\hat{\epsilon}_+}{2}; \frac{\ell}{N} ; \frac{t}{2} \right)  \, H_1\left(\frac{\hat{\epsilon}_+}{2}; -\frac{\ell}{N} ; \frac{t}{2} \right) \, \left(\cos^2 \pi \hat \epsilon_+ -\cot^2 \tfrac{\pi \ell}{N} \, \sin^2 \pi \hat\epsilon_+\right)\label{GK3}
\end{split}
\end{equation}
for the fermions associated to the non-compact $\mathbb{C}^2/\mathbb{Z}_N$ surface. In these expressions, $\ell$ denotes the sector projected by the $\ell$-th power of the orbifold generator, while we have introduced the dressed deformation parameters $\hat{\epsilon}_\pm = \epsilon_\pm t (rm -is/R)$ with $\epsilon_\pm =\epsilon_1 \pm \epsilon_2$, and
\begin{equation}
H_1 \left(z; w ; t \right)= \frac{\theta_1\left(z+w | i t \right)}{2 \eta^3(i t) \sin\pi\left(z+w\right)} \, \prod _{m\in \mathbb{Z}, n>0}\left( 1-\frac{z^2}{|m+z+w-i t n |^{2}}\right) \,.
\end{equation}

Putting together the various contributions, one gets
\begin{equation}
\begin{split}
\mathscr{A}_{g,n} \big[{\textstyle{0 \atop \ell}}\big] =
& -4 \sin^2 \left( \frac{\pi \ell}{N} \right) \,\int_0^\infty \frac{dt}{t}\, \sum_{r,s\in\mathbb{Z}}
\left( \cos^2 ( \pi \hat{\epsilon}_+) - \cot \left(\frac{\pi \ell}{N} \right)\, \sin^2 (\pi \hat{\epsilon}_+) \right)\, P_r (1/m) \, P_s (R)\, \mathbb{Z}_{K3} \big[{\textstyle{0 \atop \ell}}\big] 
\\
& \times \frac{\pi^2 (\epsilon_--\epsilon_+)(\epsilon_- +\epsilon_+)}{\sin \pi(\hat{\epsilon}_- - \hat{\epsilon}_+) \, \sin \pi (\hat{\epsilon}_- +\hat{\epsilon}_+)} \, \left[H_1\left(\frac{\hat{\epsilon}_-}{2}; 0; \frac{t}{2} \right) \right]^2
\, \frac{H_1\left(\frac{\hat{\epsilon}_+}{2}; \frac{\ell}{N}; \frac{t}{2} \right) \, H_1\left(\frac{\hat{\epsilon}_+}{2}; -\frac{\ell}{N}; \frac{t}{2} \right)}{H_1\left(\frac{\hat{\epsilon}_- -\hat{\epsilon}_+}{2}; 0 ; \frac{t}{2} \right) \, H_1\left(\frac{\hat{\epsilon}_- +\hat{\epsilon}_+}{2}; 0 ; \frac{t}{2} \right)} \,,
\end{split}
\end{equation}
where $ \mathbb{Z}_{K3} \big[{\textstyle{0 \atop \ell}}\big] $ is the standard contribution of the (non-compact) K3 bosons in the $\ell$ projected sector,
\begin{equation}
 \mathbb{Z}_{K3}  \big[{\textstyle{0 \atop \ell}}\big] = \begin{cases} \,  t^{-2} & \qquad {\rm for} \,\,   \ell=0  \,,
 \\  
 e^{2 \pi i r \ell /N} \,\, \frac{2 \eta^3 \sin\left(\frac{\pi \ell}{N}\right)}{\theta_1 \left(\frac{\ell}{N} | i t \right)} \,\,  \frac{2 \eta^3 \sin\left(-\frac{\pi \ell}{N}\right)}{\theta_1 \left(-\frac{\ell}{N} | i t \right)}\,, &\qquad  {\rm for} \,\,   \ell\not=0\,.\end{cases}
\end{equation}

Note that in this two-parameter case, the contributions of the twisted world-sheet fermions and bosons do not cancel any longer, and the infinite tower of string modes now contribute to the amplitude. 
As a result, the field theory limit we are interested in should decouple both the string oscillators and the Kaluza-Klein modes along the Scherk-Schwarz direction. This is achieved by taking  $t\to \infty$  in the theta and $H_1$ function\footnote{Remember that, in the Schwinger representation of the annulus amplitude, the $t$ modulus is proportional to $1/\alpha '$ and thus $t\to \infty$ indeed corresponds to the limit $\alpha ' \to 0$ where all massive string excitations effectively decouple.} and  $N\to \infty$  in the momentum sum. One thus finds
\begin{equation}
\begin{split}
\mathscr{F} (\epsilon_+ , \epsilon_-) &= \lim_{t,N\to \infty} \frac{1}{N} \sum_{\ell =0}^{N-1}\, \mathscr{A}_{g,n} \big[{\textstyle{0 \atop \ell}}\big] 
\\
&= \tfrac{1}{4}\pi^2 (\epsilon_- - \epsilon_+) (\epsilon_- + \epsilon_+) \, \left[ -2 F^V (\epsilon_+ , \epsilon_- ; 0 ) + 
F^H (\epsilon_+ , \epsilon_- ; m ) + F^H (\epsilon_+ , \epsilon_- ; -m ) \right]\,,
\end{split}
\end{equation}
with 
\begin{equation}
\begin{split}
F^V (\epsilon_+ , \epsilon_- ; 0 ) &= \int_0^\infty \frac{dt}{t}\, \sum_{s\in \mathbb{Z}}  \frac{1}{\sin(\pi (\hat\epsilon_- - \hat \epsilon_+ ))}\,  
\frac{1}{\sin(\pi (\hat\epsilon_- + \hat \epsilon_+ ))}\, 
\cos (2 \pi \hat\epsilon_+ ) \, e^{- \pi t (s/R)^2}
\,,
\\
F^H (\epsilon_+ , \epsilon_- ; \pm m) &= \int_0^\infty \frac{dt}{t}\, \sum_{s\in \mathbb{Z}}  \frac{1}{\sin(\pi (\hat\epsilon_- - \hat \epsilon_+ ))}\,  
\frac{1}{\sin(\pi (\hat\epsilon_- + \hat \epsilon_+ ))}\, 
e^{-\pi t (m^2 + (s/R)^2)}\,.
\end{split} \label{twoparints}
\end{equation}

\subsection{Evaluating the integrals}

Although the full amplitude $\mathscr{F} (\epsilon_+ , \epsilon_- )$ is finite, each individual contribution $F^{H,V} (\epsilon_+ , \epsilon_- ; \mu)$ is not. The integrals \eqref{twoparints} are in fact divergent in the UV $(t\to 0)$ and thus need a proper regularisation. To this end,  we perform the same change of variable as in eq. \eqref{changevar}, we sum the geometric series over the Kaluza-Klein momenta $s$, we expand 
\begin{equation}
\begin{split}
\frac{1}{\sinh (2 R \epsilon_1 t)\, \sinh (2 R \epsilon_2 t )} &= 4\, e^{ -2 \epsilon_+ R t}\,  \sum_{g_1 , g_2=0}^\infty \frac{B_{g_1}\, B_{g_2}}{g_1 ! \, g_2 !} \, (-4 R\epsilon_1 t)^{g_1-1}\, (-4 R \epsilon_2 t )^{g_2-1} 
\\
&=4\, e^{ 2 \epsilon_+ R t}\,  \sum_{g_1 , g_2=0}^\infty \frac{B_{g_1}\, B_{g_2}}{g_1 ! \, g_2 !} \, (4 R\epsilon_1 t)^{g_1-1}\, (4 R \epsilon_2 t )^{g_2-1}
\,,
\\
\frac{\cosh(2 R t (\epsilon_1 + \epsilon_2)) }{\sinh (2 R \epsilon_1 t)\, \sinh (2 R \epsilon_2 t )} &= 2 \sum_{g_1 , g_2 =0}^\infty \frac{B_{g_1}\, B_{g_2}}{g_1 ! \, g_2 !}  \left( 1 + (-1)^{g_1 + g_2 }\right) (4 R \epsilon_1 t )^{g_1 -1}\, (4 R \epsilon_2 t )^{g_2 -1}\,,
\end{split}
\end{equation}
and we deform the integration domain into the Hankel contour, as in Section \ref{evint}. Following these steps, one gets the coefficients
\begin{equation}
\begin{split}
F^H_{g_1 , g_2}  (m) &=  4  \, \frac{B_{g_1}\, B_{g_2}}{g_1 ! \, g_2 !} (g_1 + g_2 -3)!\, (4 R)^{g_1-1}\, (4 R  )^{g_2-1}\, 
\\
&\qquad \times \left[ \zeta (g_1 + g_2 -2; -2 \epsilon_+ R -i m R ) + (-1)^{g_1+g_2}\,  \zeta (g_1 +g_2 -2;1+2 \epsilon_+ R +im R) \right]\,,
\end{split}
\end{equation}
for the massive hypermultiplet, and
\begin{equation}
\begin{split}
F^V_{g_1 , g_2}  (0) &= \lim_{\mu\to 0} 2 \, \frac{B_{g_1}\, B_{g_2}}{g_1 ! \, g_2 !} (g_1 + g_2 -3)!\, \left( 1 +(-1)^{g_1 + g_2} \right) \, (4 R)^{g_1-1}\, (4 R  )^{g_2-1}\, 
\\
&\qquad \times \left[ \zeta (g_1 + g_2 -2;-i\mu R ) + \zeta (g_1 +g_2 -2;1+i\mu R)\right]\,,
\end{split}
\end{equation}
for the vector multiplet, in the double  series expansions
\begin{equation}
F^A (\epsilon_+ , \epsilon_- ; \mu ) = \sum_{g_1 , g_2 =0}^\infty F^A_{g_1 , g_2 } (\mu )  \,\epsilon_1^{g_1 -1} \, \epsilon_2^{g_2-1} \,, \qquad \qquad (A=V,H)
\,.
\end{equation}
These expressions are regular for $g_1 + g_2 > 2$, but need a proper analysis for $g_1 + g_2 \le 2$. Following a similar procedure as in Section \ref{evint}, one finds for the one-loop contribution with $g_1 + g_2 =2$ 
\begin{equation}
F^H_{2,0} (m)= F^H_{0,2}  (m) = \tfrac{1}{3} F^H_{1,1}  (m) =
-\tfrac{1}{3} \, \log \left( 2i \, \sinh \left(  2\pi i \epsilon_+ R -\pi m R\right) \right)\,,
\end{equation}
and
\begin{equation}
F^V_{2,0}  (0)= F^V_{0,2} (0) = \tfrac{1}{3} F^V_{1,1} (0) = -\tfrac{1}{3}\,  \log \, (\varLambda R ) \,.
\end{equation}
Similarly, one could derive the contributions with $g_1 + g_2 < 2$, however these do not carry any physical information in the rigid $\mathcal{N} = 2^*$ theory, and thus we do not dwell with them here.

Since for integer $n$,
\begin{equation}
\zeta (n;x)+(-1)^n \zeta (n;1-x) = \frac{(2\pi i)^n}{(n-1)!} {\rm Li}_{1-n} (e^{-2\pi i x})\,,
\end{equation}
the terms with $g_1 + g_2 >2$ admit the alternative representation
\begin{equation}
\begin{split}
F^H_{g_1 , g_2}  (m) &= 4\,  \frac{B_{g_1}}{g_1 !}\,  \frac{B_{g_2}}{g_2 !} \, (8\pi i R )^{g_1 -1} \, (8\pi i R )^{g_2-1} \, {\rm Li}_{3-g_1 -g_2} \left( e^{-2\pi mR + 4\pi i  R \epsilon_+}\right)\,,
\\
F^V_{g_1 , g_2}  (0) &= 2 \,  \frac{B_{g_1}}{g_1 !}\,  \frac{B_{g_2}}{g_2 !} \, (8\pi i R )^{g_1 -1} \, (8 \pi i R )^{g_2-1} (1+(-1)^{g_1+g_2})\, \zeta (3-g_1 -g_2)\,.
\end{split}\label{polylogrep}
\end{equation}

\subsection{The perturbative free energy}

We are now in the position to compare our results with the field theory computation of \cite{Nekrasov:2002qd, Nekrasov:2003rj}. To this end we redefine $2\pi R \to R$ and $4 i \epsilon \to \epsilon$, and get
\begin{equation}
\begin{split}
\left. \frac{\mathscr{F} (\epsilon_+ , \epsilon_- ) }{ -\pi^2 \epsilon_1 \epsilon_2 } \right|_{2\pi R \to R \atop 4 i \epsilon_i \to \epsilon_i}
&= - \tfrac{1}{3}\, \left( \frac{\epsilon_1}{\epsilon_2} + 3  + \frac{\epsilon_2}{\epsilon_1} \right)\, \left[  \log \left(2 \, \sinh \left(\tfrac{\epsilon_+ R}{4}  - \tfrac{mR}{2}  \right) \right)+\log \left(2  \, \sinh \left(\tfrac{\epsilon_+ R}{4}  + \tfrac{mR }{2} \right) \right) -  2\log \, (\varLambda R )\right] 
\\
& \quad-  4\sum_{g_1 , g_2 =0 \atop g_1 + g_2 >2}^\infty \frac{B_{g_1}}{g_1 !}\,  \frac{B_{g_2}}{g_2 !} \, ( R \epsilon_1)^{g_1 -1} \, ( R \epsilon_2)^{g_2-1} \, 
\Big[ (1+(-1)^{g_1+g_2} )\, \zeta (3-g_1 -g_2)
\\
&\qquad\qquad\qquad -  {\rm Li}_{3-g_1 -g_2} \left( e^{- mR + \frac{1}{2} R \epsilon_+}\right) - {\rm Li}_{3-g_1 -g_2} \left( e^{mR + \frac{1}{2} R \epsilon_+}\right) \Big]\,.
\end{split}
\end{equation}
Aside from irrelevant terms which depend on the regularisation scheme and have no physical relevance, this expression matches precisely the field theory result \cite{Nekrasov:2003rj}. This is perhaps more transparent if one uses the alternative representations
\begin{equation}
\begin{split}
F ^H(\epsilon_+ , \epsilon_- ; m) &=  4\, \sum_{n=1}^\infty \frac{1}{n}\, \frac{e^{- n mR  +\frac{1}{2} n R \epsilon_+ }}{(e^{n R \epsilon_1 }-1)(e^{n R\epsilon_2 }-1)} 
\\
&= 4\,  \log \, \prod_{k,\ell =1}^\infty \left( 1 - e ^{-mR - (k-\frac{1}{2}) R \epsilon_1 + (\ell - \frac{1}{2}) R \epsilon_2 }\right)\,,
\end{split}
\end{equation}
and
\begin{equation}
\begin{split}
F ^V(\epsilon_+ , \epsilon_- ; 0) &=  2\,  \sum_{n=1}^\infty \frac{1}{n}\, \frac{1 + e^{ n R \epsilon_+ }}{(e^{n R\epsilon_1 }-1)(e^{n R \epsilon_2 }-1)}
\\
&=  2\,  \log \, \prod_{k,\ell =1}^\infty \left( 1 - e^{- k R \epsilon_1  + (\ell -1) R\epsilon_2 }\right) \left( 1-e^{- (k-1) R\epsilon_1  + \ell R \epsilon_2 }\right)  \,.
\end{split}
\end{equation}

\subsection{The genus expansion}

The previous expressions \eqref{polylogrep}, although rather simple, do not provide an explicit Taylor-series expansion in the parameters $\epsilon_\pm$. In order to make more transparent the relation of the free energy with the topological amplitude \eqref{ampli2parameters} and/or the genus-$g$ partition function of the putative topological string, 
\begin{equation}
\frac{\mathscr{F} (\epsilon_+ , \epsilon_- )}{-\pi^2 \epsilon_1 \epsilon_2} = \sum_{g>0} \sum_{n\ge 0} \mathscr{F}_{g,n} (m)\, \epsilon_-^{2g-2} \, \epsilon_+^{2n}
= \sum_{g>0} \sum_{n\ge 0} \left[ -2 \mathscr{F}^V_{g,n} (0) + \mathscr{F}^H_{g,n} (m) \right] \, \epsilon_-^{2g-2} \, \epsilon_+^{2n}\,,
\end{equation}
it is convenient to express the deformation parameters $\epsilon_1$ and $\epsilon_2$ in terms of $\epsilon_\pm$, $\epsilon_{1,2} = \frac{1}{2} (\epsilon_+ \pm \epsilon_- )$. With this re-writing, powers of $\epsilon_-$ clearly count the genus expansion of the topological free energy, while $\epsilon_+$ counts the insertions of the self-dual vectors. The limit $\epsilon_- \to 0$ then reproduces the free energy in the case of a single parameter deformation. 

To proceed with the $\epsilon_\pm$ expansion, we neglect subtleties with the UV divergences, so that the formal contribution of the hypermultiplet and vector multiplet to the free energy \eqref{polylogrep} are given in terms of the (formal) double-series expansion
\begin{equation}
\frac{\mathscr{F} (\epsilon_+ , \epsilon_- )}{-\pi^2 \epsilon_1 \epsilon_2}  = \sum_{g_1 , g_2 =0}^\infty c (g_1 , g_2 )\, \epsilon_1^{g_1-1} \epsilon_2^{g_2-1}\,,
\end{equation}
where the coefficient is symmetric in $g_1$ and $g_2$, and may depend on mass deformation $m$ and the radius of the fifth dimension $R$ and, eventually also on $\epsilon_+$ in the case of the hypermultiplet. It is convenient to write
\begin{equation}
\frac{\mathscr{F} (\epsilon_+ , \epsilon_- )}{-\pi^2 \epsilon_1 \epsilon_2}  = \sum_{g_2 =0}^\infty \sum_{g_1 =0}^{g_2} c (g_1 , g_2 ) \, (\epsilon_1 \epsilon_2 )^{g_1-1} \left( \epsilon_1^{g_2-g_1} + \epsilon_2^{g_2-g_1}\right)
-\sum_{g=0}^\infty c (g,g)\, (\epsilon_1 \epsilon_2 )^{g-1} \,,
\end{equation}
so that, after binomial expansion and standard manipulation of multiple sums, one gets
\begin{equation}
\sum_{g\ge 0} c (g,g)\, (\epsilon_1 \epsilon_2 )^{g-1} = \sum_{g\ge 0} \sum_{n\ge 0} \frac{c (g+n , g+n )}{2^{2g+2n-2}}\, \left( {g+n-1 \atop n}\right)\, (-1)^{g-1}\, \epsilon_-^{2g-2}\, \epsilon_+^{2n}\,,
\end{equation}
and
\begin{equation}
\sum_{g_2 =0}^\infty \sum_{g_1 =0}^{g_2}  c (g_1 , g_2 ) \, (\epsilon_1 \epsilon_2 )^{g_1-1} \left( \epsilon_1^{g_2-g_1} + \epsilon_2^{g_2-g_1}\right)  =  \sum_{g\ge 0} \sum_{n\ge 0} \sum_{g_1=0}^{[g+n/2]} \frac{2\, c (g_1 , n+2g-g_1)}{2^{n+2g-2}}
\varPhi_{g,g_1} (n)\,\epsilon_+^n \, \epsilon_-^{2g-2}\,.
\end{equation}
where we have introduced the symbol
\begin{equation}
\varPhi_{g,g_1} (n) = \sum_{\ell=0} ^{g_1-1}\left( {n+2g-2g_1\atop n-2\ell}\right)\, \left( {g_1 -1 \atop \ell}\right) \, (-1)^{\ell +g_1+1}\,,
\end{equation}
such that $\varPhi_{g,g_1} (0) = (-1)^{g_1+1}$.

Using standard manipulations of multiple series, properties of Bernoulli numbers  and the following identity for the polylogarithms
\begin{equation}
\textrm{Li}_{-2n-a} \left(e^{-mR + \frac{1}{2} R \epsilon_+} \right) + \textrm{Li}_{-2n-a} \left(e^{mR + \frac{1}{2} R \epsilon_+} \right) = 2 \sum_{k=0}^\infty \left( \frac{R \epsilon_+}{2}\right)^{2k+1-a} \frac{\textrm{Li}_{-2n-2k-1} \left(e^{-mR}\right)}{(2k+1-a)!}\,,
\end{equation}
valid for non-negative integer $n$ and $a=0,1$, one gets
\begin{equation}
\begin{split}
\mathscr{F}^V_{g,n} (0)
&=-4\, \Biggl[ \left( \frac{B_{g+n}}{(g+n)!}\right)^2 \left( {g+n-1\atop n}\right) (-1)^{g+1} 
\\
&\qquad\qquad - 2 \sum_{g_1=0}^{n+g} \frac{B_{g_1}}{g_1 !} \frac{B_{2n+2g-g_1}}{(2n+2g-g_1 )!} \, \varPhi_{g,g_1} (2n) \Biggr]  \left( \frac{R}{2}\right)^{2n + 2g -2}\, \zeta (3-2g-2n)\, ,
\end{split}\label{twoVpm}
\end{equation}
for the contribution of the vector multiplet, and
\begin{equation}
\begin{split}
\mathscr{F}^H_{g,n} (m) 
&= -8\, \textrm{Li}_{3-2g-2n} \left(e^{-mR}\right)\, \left(\frac{R}{2}\right)^{2g+2n-2}\, \sum_{k=0}^\infty \Biggl[
\left( \frac{B_{g+n-k}}{(g+n-k)!}\right)^2 \frac{(g+n-k-1)!}{(g-1)!\, (n-k)!\, (2k)!} \, (-1)^{g+1}
\\
&\quad -2 \sum_{g_1 =0}^{g+n-k} \frac{B_{g_1}}{g_1 !} \frac{B_{2g+2n-2k-g_1}}{(2g+2n-2k-g_1)!} \frac{\varPhi_{g,g_1} (2n-2k )}{(2k)!}
\\
&\quad - \theta (n-1) \, \frac{B_{2g+2n-2k-2}}{(2g+2n-2k-2)!}\, \frac{\varPhi_{g,1} (2n-2k-1)}{(2k+1)!} \Biggr]
\end{split}\label{twoHpm}
\end{equation}
for the contribution of the hypermultiplet, with $\theta (x)$ the Heaviside function. In these expressions we have been a bit cavalier with the range of the sum over $k$. It is understood that the maximum value $k$ can assume is when the argument of any factorial becomes negative. Alternatively, we can analytically continue the above expressions in terms of Gamma functions and the finite range of the sum is automatically taken care by the poles of the Gamma function at non-positive integers.

As expected, in the special case $\epsilon_1 = - \epsilon_2 = \hbar$ of a one-parameter deformation, these expressions reproduce eq. \eqref{5dfreeen}. In fact, now only the term with $n=0$ survives and the remarkable identity
\begin{equation}
\left( \frac{B_g}{g!}\right)^2 (-1)^g - 2 \sum_{g_1=0}^{g} \frac{B_{g_1}}{g_1 !}\, \frac{B_{2g-g_1}}{(2g-g_1)!} \, (-1)^{g_1} = \frac{B_{2g}}{2g (2g-2)!}\,,
\end{equation}
satisfied by the Bernoulli numbers, together with $\varPhi_{g,g_1} (-2k) = (-1)^{g_1+1} \delta_{k,0}$, for $k\ge 0$, leads to the desired result.

\subsection{The four-dimensional limit}

As usual, the four-dimensional free-energy is obtained by taking $R\to 0$. In this limit the Kaluza-Klein states along the fifth direction decouple and one is left with an effective four-dimensional $\mathscr{N}=2^*$ theory. Using eq. \eqref{limitpolylog}, one readily finds 
\begin{equation}
\begin{split}
\frac{\mathscr{F} (\epsilon_1 , \epsilon_2)}{-\pi^2 \epsilon_1 \epsilon_2} &= \tfrac{1}{3} \left(\frac{\epsilon_1}{\epsilon_2} + 3 + \frac{\epsilon_2}{\epsilon_1} \right) \, \left[ \log \left( \frac{\varLambda^2}{m^2 - \frac{1}{4} \epsilon_+^2}\right) + \textrm{const}\right]
\\
&\quad + 4\,\sum_{{g_1 , g_2 =0 \atop g_1 + g_2 >2}}^\infty \frac{B_{g_1}}{g_1 !} \frac{B_{g_2}}{g_2 !} \, (g_1 + g_2 -3)! \, \epsilon_1^{g_1-1} \epsilon_2^{g_2-1}\, \frac{ \left( \frac{1}{2} \epsilon_+ +m \right)^{g_1 + g_2 -2} + \left(\frac{1}{2} \epsilon_+ - m \right)^{g_1 + g_2 -2}}{\left( m^2 - \frac{1}{4} \epsilon_+^2 \right)^{g_1 +  g_2 - 2}}\,.
\end{split}
\end{equation}
Using eqs. \eqref{twoVpm} and \eqref{twoHpm} one could have obtained the alternative expansion in powers of $\epsilon_\pm$.  
Notice that, as expected, for the choice $\epsilon_1 = - \epsilon_2 = \hbar$ one recovers eq. \eqref{4doneparfree}.

\section{The non-Abelian extension}\label{nonabelian}

The previous results can be straightfowardly generalised to the non-Abelian $\textrm{U}(M)$ $\mathscr{N}=2^*$ gauge theory. In the D-brane construction this configuration is simply obtained by piling-up $M$ D-branes. Since the freely-acting orbifold of Section \ref{SSString} does not act on the Chan-Paton factors, the annulus amplitude for the non-Abelian case is obtained from eq. \eqref{annulusSSS} by simply multiplying it by $M^2$. As a result, the free energies with one or two parameter deformations are also multiplied by $M^2$. 

The Coulomb branch is obtained by introducing suitable Wilson lines along the compact directions which break $\textrm{U}(M) \to \textrm{U} (1)^M$ \cite{Bianchi:1990tb,Angelantonj:2002ct}. In the string realisation described in Section \ref{SSString} we have then the option of deforming the theory by  introducing either Wilson lines  $a_i^m$ along the Scherk-Schwarz circle $S^1_m $, or Wilson lines $a^R_i$ along the spectator circle $S^1 _R$, or  both $a_i^m$ and $a^R_i$. The annulus amplitude  reads in general
\begin{equation}
\mathscr{A}_{\textrm{Coulomb}} = \sum_{i,j=1}^M \frac{1}{N} \left[ \sum_{\ell =0}^{N-1} \rho \left[ {0\atop \ell}\right] \, \sum_{r\in\mathbb{Z}} e^{2 i \pi r \ell /N} \, P_{r + a^m_i - a^m_j} (1/m) \right]\, \sum_{s\in \mathbb{Z}} P_{s+a_i^R - a_j^R} (R)\,.
\end{equation}
To describe the field theory limit, it is convenient to introduce the complex Wilson line
\begin{equation}
a_i = m\, a^{m}_i + i \, \frac{a^R_i}{R}\,,
\end{equation}
so that the free energies read
\begin{equation}
\begin{split}
(\pi \hbar )^{-2} \, \mathscr{F} (\hbar ) &= \sum_{i,j=1}^M \sum_{\mu = 0,\pm m} d(\mu )\, F (\hbar ; \mu + a_i - a_j )\,,
\\
\frac{\mathscr{F} (\epsilon_+ , \epsilon_- )}{-\pi^2 \epsilon_1 \epsilon_2} &= \tfrac{1}{4} \pi^2 (\epsilon_- - \epsilon_+)( \epsilon_- + \epsilon_+ ) \sum_{i,j=1}^M
 \left[ -2 F^V (\epsilon_+ , \epsilon_- ; a_i - a_j  ) \right.
 \\
 &\qquad \qquad\qquad \qquad \left.
+ F^H (\epsilon_+ , \epsilon_- ; m +a_i - a_j) + F^H (\epsilon_+ , \epsilon_- ; -m  +a_i - a_j) \right]\,,
\end{split}
\end{equation}
with the $F$'s given in eqs. \eqref{freeen} and \eqref{polylogrep} for the one parameter background and the two parameter background, respectively. 

These results reproduce those of \cite{Florakis:2015ied} and extend them to the case of an $\varOmega$ background with two parameters.

\section{Conclusions}\label{concls}

In this work, we presented a realisation of an Abelian $\mathscr{ N}=2^*$ theory and its radius deformation in five dimensions in a D-brane set-up, whereby a freely acting orbifold breaks spontaneously $\mathscr{N}=4$ supersymmetry to $\mathscr{N}=2$, as a Scherk-Schwarz deformation. We then computed a series of generalised topological amplitudes on the world-sheet annulus, involving four gravitini, $2g-2$ anti-self-dual graviphotons and $n$ self-dual gauge fields belonging to the multiplet of the D5-brane coupling modulus. We checked that in the field theory limit these amplitudes reproduce the perturbative part of the Nekrasov partition function in the $\varOmega$ background~\cite{Nekrasov:2003rj}, as a power expansion of the two deformation parameters, in agreement with the proposal of Ref.~\cite{Antoniadis:2013epe}. We also discussed the case of  non-Abelian unitary gauge grups, generalising the work of Ref.~\cite{Florakis:2015ied}.

Our results can be in principle extended to include non-perturbative contributions by using the open string description of D-brane instantons, following the formalism of Refs.~\cite{Billo:2006jm,Antoniadis:2013mna, Moskovic:2016xhn}. Since the $\textrm{U}(1)$ case presents several peculiarities despite its simplicity, we plan to come back to it in the future.

\section*{Acknowledgements}

We thank Francesco Fucito, Domenico Orlando, Rubik Poghossian and Susanne Reffert for stimulating discussions. 
C.A. is grateful to the Albert Einstein Center for Fundamental Physics of the Institute for Theoretical Physics of Bern University and the Theory Department of CERN for hospitality during various stages of this work. M.S. is grateful to the Albert Einstein Center for Fundamental Physics of the Institute for Theoretical Physics of Bern University and the Physics Department of Turin University for hospitality during various stages of this work. The work of C.A. has been partially supported by the Compagnia di San Paolo contract ``Modern Application in String Theory'' (MAST) TO-Call3-2012-0088 and by the  INFN Special Project "Gauge theories, Supergravity and String Theory".


\bibliographystyle{unsrt}

\end{document}